# On-chip rewritable phase-change metasurface for programmable diffractive deep neural networks

## Sanaz Zarei

Department of Electrical Engineering  Sharif University of Technology, Tehran, Iran

szarei@sharif.edu

*Abstract*—Photonic neural networks capable of rapid programming are indispensable to realize many functionalities. Phase change technology can provide nonvolatile programmability in photonic neural networks. Integrating direct laser writing technique with phase change material (PCM) can potentially enable programming and in-memory computing for on-chip photonic neural networks. $Sb_2Se_3$ is a newly introduced ultralow-loss phase change material with a large refractive index contrast over the telecommunication transmission band. Compact, low-loss, rewritable, and nonvolatile on-chip phase-change metasurfaces can be created by using direct laser writing on a $Sb_2Se_3$ thin film. Here, by cascading multiple layers of on-chip phase-change metasurfaces, an ultra-compact on-chip programmable diffractive deep neural network is demonstrated at the wavelength of 1.55µm and benchmarked on two machine learning tasks of pattern recognition and MNIST (Modified National Institute of Standards and Technology) handwritten digits classification and accuracies comparable to the state of the art are achieved. The proposed on-chip programmable diffractive deep neural network is also advantageous in terms of power consumption because of the ultralow-loss of the $Sb_2Se_3$ and its nonvolatility which requires no constant power supply to maintain its programmed state.

*Keywords*—on-chip programmable diffractive deep neural network, phase-change metasurface, $Sb_2Se_3$, direct laser writing, machine learning

## 1. INTRODUCTION

Recently, direct laser writing on $Sb_2Se_3$ phase-change thin films has been exploited as a simple, low-cost, and fast approach to create or reprogram photonic integrated circuits (PICs) [1]. Various photonic components consisting of waveguides, gratings, ring resonators, couplers, crossings, and interferometers were created in one writing step, without the need for additional fabrication processes [1]. Furthermore, writing imperfections (errors) can be easily corrected by locally adjusting each element using erasing and restoring [1]. Before this, Delany et al. employed the combination of large refractive index contrast and ultralow losses of $Sb_2Se_3$ and demonstrated the feasibility of a laser-programmed multiport router based on writing nonvolatile patterns of weakly scattering perturbations onto a multimode interference device [2]. Blundell et al. continued this work and could also establish the ability to strongly increase the effect per pixel/unit length by increasing the $Sb_2Se_3$ film thickness up to 100nm in their very recent article [3]. Wu et al. took the advantage of direct laser writing of phase patterns optimized by the inverse design technique to achieve reconfigurability in programmable phase-change photonic devices [4]. They could demonstrate a photonic device that can be reconfigured from a wavelength-division demultiplexer to a mode-division demultiplexer [4].

Photonic neural networks (PNNs) may be the next-generation computing platforms due to their advantages of low power consumption, high parallelism, and light-speed signal processing. On-chip diffractive optical neural networks are one of many implementations of PNNs [5-7], in which an on-chip high-contrast transmit array (HCTA) metasurface [8] functions as an optical neural network hidden layer. Inspired by the previously presented on-chip diffractive optical neural networks [5-7] and leveraged on the technique of direct laser writing on $Sb_2Se_3$ thin film, an on-chip programmable diffractive deep neural network is presented in this short communication for image classification.

## 2. PHASE-CHANGE METASURFACE

In this study, the on-chip programmable diffractive deep neural network consists of multiple phase-change metasurfaces, each of which performs as a neural network layer. The phase-change metasurface is a one-dimensional amorphous $Sb_2Se_3$ ($aSb_2Se_3$) rods array in the crystalline $Sb_2Se_3$ ($cSb_2Se_3$) thin film as is shown in Fig. 1(a). Analogous to previous demonstrations [1-4], the metasurface can be written (or rewritten) using direct laser writing. The lattice constant of the metasurface is 500nm. The $Sb_2Se_3$ film is 30nm thick and it is coated with a $SiO_2$ capping layer of 200nm thickness [1]. The $SiO_2$ layer is for protection and oxidation inhibition of the $Sb_2Se_3$ layer. Underneath the $Sb_2Se_3$, there is a $Si_3N_4$ film with 330nm thickness on a standard oxidized silicon substrate [1]. A single neuron (or meta-atom) is formed by a single ($aSb_2Se_3$) rod. The geometrical parameters of an ($aSb_2Se_3$) rod can be learnable parameters. As is shown in figures 1(b) and 1(c), by adjusting the geometrical parameters of an ($aSb_2Se_3$) rod (i.e. the ($aSb_2Se_3$) rod length and width), the transmission phase shift and amplitude of a single neuron for TE-polarized guided wave can be tuned. By fixing the (aSb$_2$Se$_3$) rod width to 400nm and altering the rod length from 300nm to 4μm, more than π/2 phase shift is attained, while the transmission amplitude is very near to 1 (Fig. 1(d)). Therefore, in the on-chip programmable diffractive deep neural network, the rod lengths of meta-atoms (neurons) are chosen as the learnable parameters, which collectively adjust the amplitude and phase profile of the output wavefront. The plots in Fig. 1 are generated using the commercial software package Lumerical FDTD. Throughout simulations, the fundamental TE mode is selected for excitation, and the x-axis is chosen as the injection axis. The operation wavelength is set to be 1.55μm.

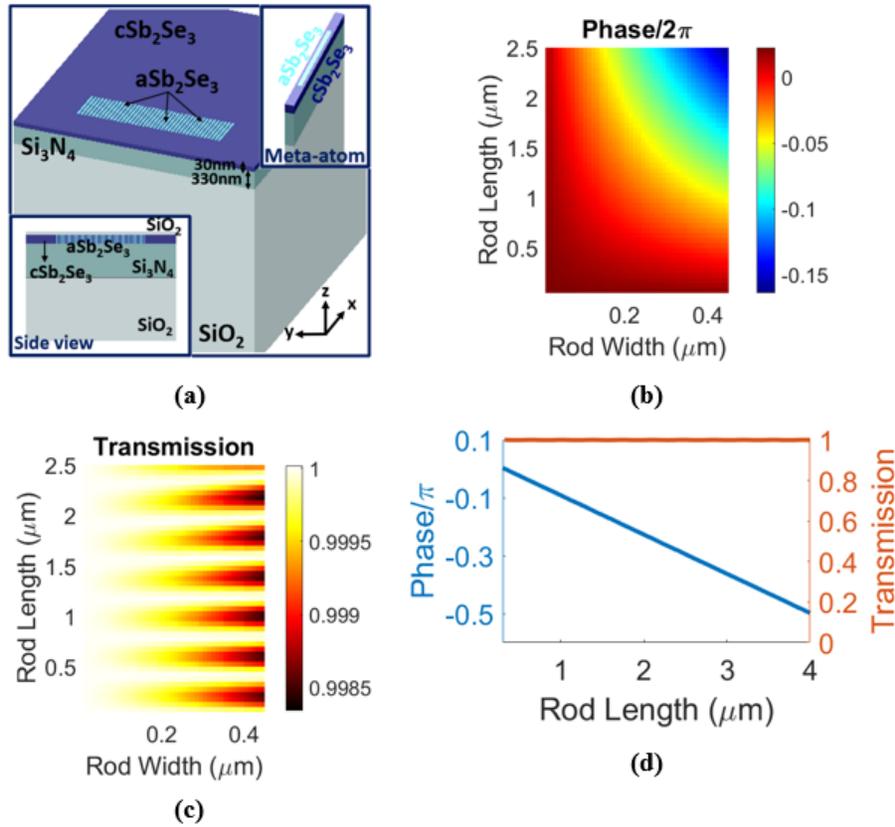

Fig. 1. (a) Schematic of (aSb$_2$Se$_3$) rods array in the (cSb$_2$Se$_3$) thin film. The insets show the side view and a meta-atom of the structure. The electric field transmission phase (b) and amplitude (c) of the TE-polarized guided wave for a meta-atom versus the (aSb$_2$Se$_3$) rod length and width. (d) Electric field transmission phase and amplitude versus (aSb$_2$Se$_3$) rod length, fixing the rod width at 400nm. These plots are generated by the commercial software package Lumerical FDTD. The operation wavelength is set to be 1.55μm.

## 3. ON-CHIP PROGRAMMABLE DIFFRACTIVE DEEP NEURAL NETWORK

Like artificial deep neural networks, an on-chip programmable diffractive deep neural network is comprised of one input layer, multiple hidden layers, and one output layer. Each hidden layer of the network is a phase-change metasurface that consists of many meta-atoms (neurons) arranged linearly. Each meta-atom is a weight element that connects to the meta-atoms on adjacent layers through diffraction and interference of light [5]. The input images are preprocessed and converted to a vector and then encoded into the amplitude of the input light at the input layer. The output layer is composed of multiple detection regions aligned in a linear configuration. The network is trained on the training dataset using an error-backpropagation algorithm based on the adjoint gradient method described in [9-10]. After numerical training, the inference performance of the designed network is numerically tested using the test dataset and subsequently verified using the 2.5D variational FDTD solver of the Lumerical Mode Solution. Here, the capability of the on-chip programmable diffractive deep neural network is benchmarked on two machine learning tasks of pattern recognition for English letters X, Y, and Z and MNIST (Modified National Institute of Standards and Technology) handwritten (0-1-2) digits classification.

### 3.1. Pattern Recognition

First, the on-chip programmable diffractive deep neural network is trained for pattern recognition of English letters X, Y, and Z. The input letters are the binary letter images with 60 pixels (Fig. 2(a)). The dataset is created by amplitude flipping in random single-pixel and double-pixels in the binary letter images of X, Y, and Z [5]. This generates 5490 images, of which 4590 images are used as the training dataset and fed through the metasystem in batches of 10 during each training epoch, and the remaining 900 images are used as the test dataset.

The input patterns are reshaped from a two-dimensional 10×6 matrix to a 60-component one-dimensional vector [5] and then encoded as the amplitude of the input light. The pattern recognizer is composed of five phase-change metasurfaces, each of which consists of 60 meta-atoms (Fig. 2(b)). The 30μm-length metasurfaces are precisely aligned with 8μm separations. After light exits the fifth metasurface, it propagates 8μm until it reaches the network output layer with three linearly arranged detectors (corresponding to the letters X, Y, and Z). Fig. 2(c) illustrates the evolution of the loss and accuracy for the training set during the learning procedure. Only after 3 epochs, the metasystem can achieve 100% accuracy in English letters recognition. The blind testing accuracy of the designed metasystem is 100% over the test dataset. Verification with Lumerical 2.5D FDTD also indicates 98.8% matching with numerical testing results. For verification, 90 random patterns, 30 random patterns per letter, are selected from the test dataset. The confusion matrices for numerical testing and FDTD testing are shown in figures 2(d) and 2(e), respectively. The x-y view of electric field distribution in the network and the normalized power of three output detectors for a sample input pattern of letter Y are presented in Fig. 2(f). Table I summarizes the characteristics of the designed pattern recognizer.

*3.2. MNIST Handwritten Digits Classification*

For digit classification, the training is performed using 18623 images of handwritten digits (0-1-2) from the MNIST handwritten digits database. The training images are fed through the network in batches of 64. After the training phase, the designed network is tested on 3147 images of the test dataset. The grayscale 28×28-pixel images are down-sampled to 14×14 pixels and converted to 196-component vectors. The vectors are then encoded as the amplitude of the input light.

The schematic of the digit classifier is illustrated in Fig. 3(a). It is composed of three phase-change metasurfaces, each with 196 meta-atoms. The length of the one-dimensional metasurfaces is 98μm and they are precisely aligned with 7μm separations. After light exits the third metasurface, it propagates 7μm until it reaches the output layer of the network with three linearly aligned detectors (corresponding to the digits 0, 1, and 2). Fig. 3(b) depicts the loss and training accuracy in each epoch during the learning process. After 140 epochs, the training accuracy reaches 92.38% over the whole training dataset. The blind testing accuracy of the finally-designed metasystem is 91.86% over the test dataset. From the test dataset images that the designed network can successfully classify, 100 handwritten digit images are randomly selected for 2.5D FDTD verification. Verifications indicate that for 92 out of 100 images, similar predictions to numerical predictions are made, which means 92% matching between the two predictions. The confusion matrices for numerical testing and FDTD testing of the network are shown in figures 3(c) and 3(d), respectively. Also, the dependence of the digit classifier performance on the number of metasurfaces (hidden layers) is investigated in Fig. 3(e). As can be observed, the network blind testing accuracy grows from 86.30% for the one-layer network to 94.43% for the four-layer network and then declines to 92.50% for the five-layer network. The matching percentage between numerical testing and FDTD testing is above 90% for all the neural networks. The one-layer network has the best matching score of 98% and the five-layer network has the worst matching score of 91%. Due to the nearly similar performance of all the networks in Fig. 3(e), the structural complexity and footprint may be the main determining factors. Fig. 3(f) shows the x-y view of electric field distribution in the three-layer metasystem and the normalized power of three output detectors for a sample image of the handwritten digit 2. Table II summarizes the characteristics of the presented image classifier.

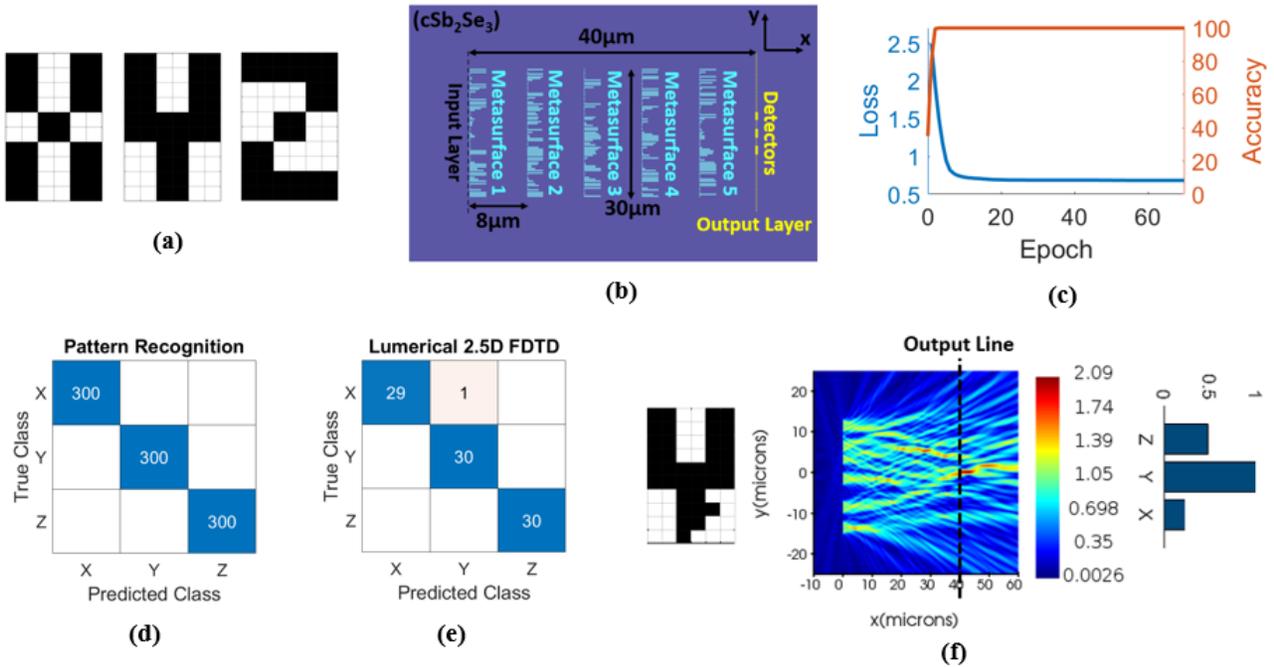

Fig. 2. (a) 10×6-pixel images of the letters X, Y and Z, (b) 2D schematic of the five-layer network trained to perform as an on-chip pattern recognizer, (c) loss and training accuracy versus epoch during the learning process, (d) the confusion matrix of the network over 900 images of the test dataset, (e) the confusion matrix of the network, computed by 2.5D FDTD of Lumerical Mode Solution, over 90 random images from the test dataset, (f) the x-y view of the electric field distribution at the middle plane of the Sb$_2$Se$_3$ thin film through the whole network and the normalized power of the three output detectors for a representative Y image.

TABLE I. SUMMARY OF THE CHARACTERISTICS FOR THE METASYSTEM TRAINED FOR PATTERN RECOGNITION.

| | |
|---|---|
| Number of metasurfaces | 5 |
| Number of meta-atoms per metasurface | 60 |
| Number of design variables | 300 |
| Length of each on-chip metasurface | 30μm |
| Distance between the input layer and the first metasurface | 0 |
| Distance between two neighboring metasurfaces | 8μm |
| Distance between the last metasurface and the output layer | 8μm |
| Total device size | 30μm×40μm |
| Number of detectors | 3 |
| Length of each detector | 2μm |
| Distance between two neighboring detectors | 2μm |
| Detectors arrangement | Linear |
| Blind testing accuracy on pattern recognition | 100% |
| Matching between numerical predictions and Lumerical 2.5D FDTD verifications | 98.8% |

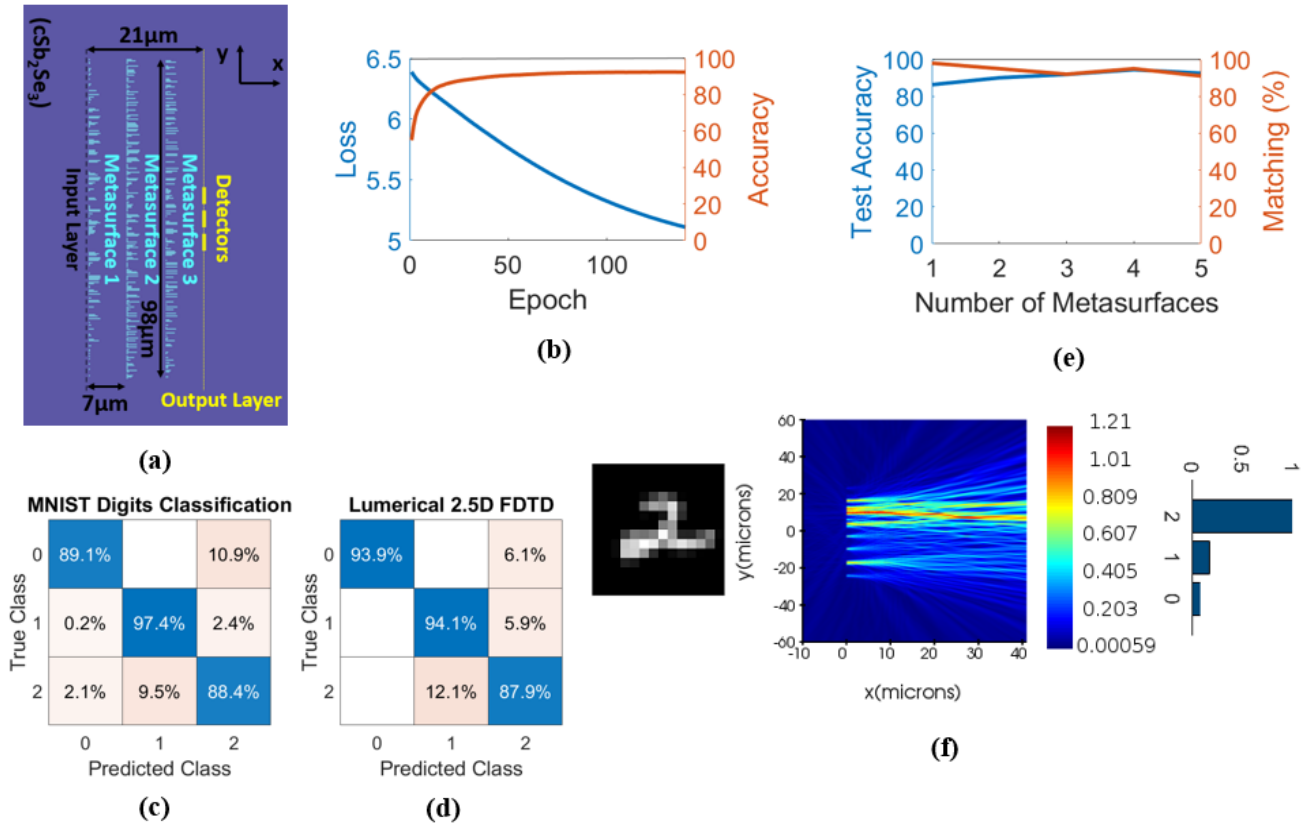

Fig. 3. (a) 2D schematic of the three-layer network trained to perform MNIST (0-1-2) handwritten digits classification, (b) loss and training accuracy versus epoch during the learning process, (c) the confusion matrix of the metasystem over 3147 images of the test dataset, (d) the confusion matrix of the metasystem, calculated by 2.5D FDTD of Lumerical Mode Solution, over 100 random images from the test dataset, (e) blind testing accuracy for metasystems with different number of metasurfaces and the associated matching percentage between numerical testing and Lumerical 2.5D FDTD verification over 100 random images from the test dataset, (f) the x-y view of the electric field distribution at the middle plane of the $Sb_2Se_3$ thin film through the metasystem and the normalized power of three output detectors for a representative image of the digit 2.

TABLE II. SUMMARY OF THE CHARACTERISTICS FOR THE METASYSTEM TRAINED AS (0-1-2) DIGIT CLASSIFIER.

| | |
|---|---|
| Number of metasurfaces | 3 |
| Number of meta-atoms per metasurface | 196 |
| Number of design variables | 588 |
| Length of each on-chip metasurface | 98µm |
| Distance between the input layer and the first metasurface | 0 |
| Distance between two neighboring metasurfaces | 7µm |
| Distance between the last metasurface and the output layer | 7µm |
| Total device size | 98µm×21µm |
| Number of detectors | 3 |
| Length of each detector | 5µm |
| Distance between two neighboring detectors | 2µm |
| Detectors arrangement | Linear |
| Blind testing accuracy on MNIST (0-1-2) digit classification | 91.86% |
| Matching between numerical predictions and Lumerical 2.5D FDTD verifications | 92% |

## 4. CONCLUSION

In conclusion, by applying the direct laser writing technique on $Sb_2Se_3$ phase change material to direct-write or rewrite phase-change metasurfaces, an ultra-compact on-chip programmable diffractive deep neural network can be created. This technique provides a simple, low-cost, and fast approach to reprogram diffractive deep neural networks. With this reconfigurable and nonvolatile metasystem, the feasibility of a variety of functionalities such as optical computing and machine learning tasks can be explored more rapidly. In this communication, numerical testing and FDTD testing results for two machine learning tasks of three-letter pattern recognition and more complicated three-handwritten-digits (from MNIST dataset) classification imply that the network can properly perform these tasks.

## FUNDING

This research did not receive any specific grant from funding agencies in the public, commercial, or not-for-profit sectors.

## COMPETING INTERESTS

The authors have no relevant financial or non-financial interests to disclose.

## DATA AVAILABILITY

The data that support the findings of this study are available from the corresponding author upon reasonable request.